\documentclass[aps,pre,amsmath,amssymb,showpacs,amsfonts]{revtex4}
\usepackage{epsfig}
\usepackage{graphicx}
\usepackage{dcolumn}
\usepackage{bm}

\begin{document}

\title{CFT Hydrodynamics: Symmetries, Exact Solutions and Gravity}

\author{Itzhak Fouxon}
\author{Yaron Oz}

\affiliation{Raymond and Beverly Sackler School of Physics and Astronomy, Tel-Aviv University, Tel-Aviv 69978, Israel}

\date{\today}

\begin{abstract}
We consider the hydrodynamics of relativistic conformal field theories at finite temperature
and its slow motions limit, where it reduces to the incompressible Navier-Stokes equations. The symmetries of the equations and their solutions are analyzed.
We construct exact solutions with finite time singularities of one-dimensional relativistic conformal hydrodynamic motions, and use them to generate multi-dimensional solutions via  special
conformal transformations. These solutions, however, are shown to have no non-trivial slow motions
limit. A simple non-equilibrium steady state in the form of a shock solution is constructed, and its inner structure is analyzed.
We demonstrate that the derivation of the gravitational dual description of conformal hydrodynamics is analogous
to the derivation of hydrodynamics equations from the Boltzmann equation.
The shock solution is shown to correspond to a domain-wall solution in gravity. We show that the solutions to the non-relativistic
incompressible Navier-Stokes equations play a special role in the construction of global solutions to gravity.

\end{abstract}


\maketitle

\section{Introduction}

The hydrodynamics of relativistic conformal field theories has attracted much attention
recently, largely in view of the AdS/CFT correspondence between gravitational theories on asymptotically
Anti-de-Sitter (AdS) spaces and CFTs \cite{Maldacena:1997re} (for a review see \cite{Aharony:1999ti}).
Hydrodynamics gives a universal description of the large-time dynamics of the theory: starting from an arbitrary
initial state of the CFT, within the correlation time $\tau_{cor}$ the system approaches the state of local thermal
equilibrium. At $t\gg \tau_{cor}$ the evolution is mainly the evolution of the parameters of the local
equilibrium described by hydrodynamics. On the basis of the AdS/CFT correspondence one then expects
that the large-time dynamics of gravity can be obtained as a dual description of
the CFT hydrodynamics \cite{Bhattacharyya:2008jc,Minwalla5}.

A complete (compressible) hydrodynamics is described by  five fields: the three velocity components, the temperature and the
particle density \cite{Frisch,Landau,Batchelor}.
In a CFT there is no locally conserved charge corresponding to the particle density \cite{Landau,JeonYaffe}.
As a result, conformal hydrodynamics is described by
only four fields: the three velocity components and the temperature.

Hydrodynamics applies
under the condition that the correlation length of the fluid $l_{cor}$ is much smaller than the characteristic scale $L$ of variations of the macroscopic fields.
In order to characterize this, one introduces
the dimensionless Knudsen number
\begin{equation}
Kn\equiv l_{cor}/L \ .
\end{equation}
Since the only dimensionfull parameter is the characteristic temperature of the fluid
$T$, one has by dimensional analysis,
\begin{eqnarray}&&
l_{cor}=(\hbar c/k_B T)G(\lambda) \ , \label{eq112}
\end{eqnarray}
where $\lambda$ denotes all the dimensionless parameters of the CFT. The function $G(\lambda)$ characterizes the CFT.

The stress-energy tensor of the CFT obeys
\begin{equation}
\partial_{\nu}T^{\mu\nu}=0,~~~~~T^{\mu}_{\mu}=0 \ .
\label{cfteq}
\end{equation}
The
equations of relativistic hydrodynamics are determined by the constitutive relation expressing $T^{\mu\nu}$ in terms of the temperature $T(x)$ and the four-velocity field $u^{\mu}(x)$ satisfying $u_{\mu}u^{\mu}=-1$. Here $u^{\mu}$ and $T$ determine the local thermal equilibrium distribution of the fluid.
The constitutive relation has the form of a series in the small parameter $Kn\ll 1$,
\begin{eqnarray}&&
T^{\mu\nu}(x)=\sum_{l=0}^{\infty}T^{\mu\nu}_l(x),\ \ T^{\mu\nu}_l\sim (Kn)^l, \label{series}
\end{eqnarray}
where $T^{\mu\nu}_l(x)$ is determined by the local values of $u^{\mu}$ and $T$ and their
derivatives of a finite order. The smallness of $T^{\mu\nu}_l$ arises because it involves either the $l-$th derivative of $u^{\mu}$ or $T$ or because it contains the corresponding power of a lower-order derivative.
Keeping only the first term in the series gives ideal hydrodynamics, within which the entropy
is conserved and the entropy density per unit volume $\sigma$ obeys a conservation law $\partial_{\mu}(\sigma u^{\mu})=0$.
Dissipative hydrodynamics arises when
one keeps the first two terms in the series.

The ideal hydrodynamics
approximation for $T^{\mu\nu}$ does not contain
the spatial derivatives of the fields. The stress-energy tensor reads (up to a multiplicative constant)
\begin{eqnarray}&&
T_{\mu\nu}= T^4\left[\eta_{\mu\nu}+4u_{\mu}u_{\nu}\right] \ ,
\label{ideal00}
\end{eqnarray}
where $\eta_{\mu\nu} = diag[-,+,+,+]$.

The dissipative hydrodynamics is obtained by keeping the $l=1$ term in the
series in Eq.~(\ref{series}). In the Landau frame
\cite{Landau,Son} the stress-energy tensor reads (up to a multiplicative constant)
\begin{eqnarray}&&
T_{\mu\nu}=T^4\left[\eta_{\mu\nu}+4u_{\mu}u_{\nu}\right]-c\eta\sigma_{\mu\nu} \ ,
\label{visc00}
\end{eqnarray}
where $\sigma_{\mu\nu}$ obeys $\sigma_{\mu\nu}u^{\nu}=0$ and is given by
\begin{eqnarray}&&
\sigma_{\mu\nu}=\left(\partial_{\mu} u_{\nu}+
\partial_{\nu} u_{\mu}+u_{\nu}u^{\rho}\partial_{\rho} u_{\mu}+
u_{\mu}u^{\rho}\partial_{\rho} u_{\nu}\right) \nonumber\\&&
-\frac{2}{3}\partial_{\alpha}u^{\alpha}
\left[\eta_{\mu\nu}+u_{\mu}u_{\nu}\right] \ . \label{str}
\end{eqnarray}
The dissipative hydrodynamics of a CFT is determined by only one kinetic coefficient -
the shear viscosity $\eta$.
The bulk viscosity $\zeta$ vanishes for the CFT, while the absence
of the particle number conservation and the use of the Landau frame allow to avoid the use of
heat conductivity, which is not an independent coefficient here \cite{JeonYaffe}. Dimensional
analysis dictates that $\eta=F(\lambda)T^3$, where $F(\lambda)$ is a function characterizing the CFT
and we again omit
a multiplicative constant in $T^{\mu\nu}$.

The hydrodynamics of relativistic conformal field theories is intrinsically relativistic as is
the microscopic dynamics.
However, it has been shown in \cite{Fouxon:2008tb},
that the limit of non-relativistic macroscopic motions of a CFT hydrodynamics
leads to the non-relativistic incompressible Euler
and Navier-Stokes equations for ideal and dissipative hydrodynamics of the CFT, respectively (see also \cite{Bhattacharyya:2008kq}). For ideal hydrodynamics the implication follows by noting that the equations (\ref{cfteq}) written in terms of $\bm v$ defined by $u^{\mu}=(\gamma, \gamma\bm v/c)$ and a variable $P$ equal to $c^2\ln T$ up to an additive constant, take the form
\begin{eqnarray}&&
\frac{1}{c^2}\left[
\frac{\partial P}{\partial t}+\frac{2c^2}{3c^2-v^2}(\bm v\cdot\nabla)P\right]=-\frac{c^2}{3c^2-v^2}\nabla\cdot\bm v,
\label{hyd3}\\&&
\frac{\partial v_i}{\partial t}+(\bm v\cdot\nabla) v_i=-\left(1-\frac{v^2}{c^2}\right)\left[\delta_{ij}-\frac{2v_iv_j}{3c^2-v^2}\right]
\nabla_jP
+\frac{(c^2-v^2) v_i(\nabla\cdot\bm v)}{3c^2-v^2} \ . \label{hyd4}
\end{eqnarray}
Now if we consider the solutions for which $\bm v$ remains finite in the limit $c\to\infty$, then the limiting field $\bm v$ obeys
\begin{eqnarray}&&
\frac{\partial \bm v}{\partial t}+(\bm v\cdot\nabla) \bm v=-\nabla P+\frac{\nabla\cdot\bm v}{3}\bm v,\ \ -\frac{1}{3}\nabla\cdot\bm v=\frac{1}{c^2}\frac{\partial P}{\partial t} \ . \label{Eul}
\end{eqnarray}
The first equation above implies that $\nabla P$ must be finite at $c\to\infty$. However it is still
possible that $P$ contains a function of time leading to a finite spatially constant divergence of velocity. The general form of non-relativistic dynamics is therefore
\begin{eqnarray}&&
\frac{\partial \bm v}{\partial t}+(\bm v\cdot\nabla) \bm v=-\nabla P-a(t)\bm v,\ \ \nabla\cdot\bm v=-3a(t) \ . \label{Eul1}
\end{eqnarray}
If we now impose the condition that $\bm v$ remains finite at large distances (which is valid in most physical
situations), then we must require that $a(t)\equiv 0$. We conclude that non-relativistic, finite at infinity,
motions of the CFT obey the incompressible Euler equations
\begin{eqnarray}&&
\frac{\partial \bm v}{\partial t}+(\bm v\cdot\nabla) \bm v=-\nabla P,\ \ \nabla\cdot\bm v=0 \ . \label{Eul2}
\end{eqnarray}
Analogous considerations hold for the viscous hydrodynamics where the limiting field $\bm v$ obeys the incompressible Navier-Stokes equations
\begin{eqnarray}&&
\frac{\partial \bm v}{\partial t}+(\bm v\cdot\nabla) \bm v=-\nabla P+\nu\nabla^2 \bm v,\ \ \nabla\cdot\bm v=0 \ . \label{NS}
\end{eqnarray}
The kinematic viscosity $\nu$ is given by
\begin{equation}
\nu=\hbar c^2 F(\lambda)/4k_BT_0 \ .
\end{equation}
$T_0$ is the main component of the
temperature, which is approximately constant in the considered limit,
\begin{eqnarray}&&
T=T_0\left[1+\frac{P}{c^2}+O\left(\frac{1}{c}\right)\right] \ ,
\end{eqnarray}
see \cite{Fouxon:2008tb}.
Thus, a relativistic conformal field theory contains the incompressible Euler and Navier-Stokes equations inside it.

The solutions to the non-relativistic incompressible Navier-Stokes equations have special importance in light of the AdS/CFT correspondence: they allow to construct an explicit approximate solutions to
the five-dimensional Einstein equations with negative cosmological constant
\begin{eqnarray}&&
R_{mn}+4g_{mn}=0,\ \ R=-20 \ , \label{u7}
\end{eqnarray}
where we use $R_{AdS}=1$.
If $\bm v$ and $P$ solve Eqs.~(\ref{NS}) and obey the condition $Kn\ll 1$, then the metric $g_0$ defined by ($\hbar=k_B=1$)
\begin{eqnarray}&&
(g_0)_{mn}dy^{m}dy^{n}=-2u_{\mu}(x^{\alpha})dx^{\mu}dr+\pi^4 T^4(x^{\alpha}) r^{-2}u_{\mu}(x^{\alpha})u_{\nu}(x^{\alpha})dx^{\mu}dx^{\nu}+r^2\eta_{\mu\nu}dx^{\mu}dx^{\nu},\label{gravityansatz01} \\&&
y=(x^{\mu}, r),\ \ u^{\mu}=\left(\frac{1}{\sqrt{1-v^2/c^2}}, \frac{\bm v/c}{\sqrt{1-v^2/c^2}}\right),\ \
T = \frac{c^2}{4\pi\nu} + \frac{P}{4\pi \nu} \ ,
\label{gravityansatz02}
\end{eqnarray}
has a value of $R_{mn}+4g_{mn}$, which is small both in $Kn$ and $v/c$ (see the details in section \ref{gravity}).

Consider a typical solution of Eq.~(\ref{NS}), which is turbulent (the majority of flows in nature are turbulent). There are two important
scales in the solution: the outer scale of turbulence  $L_O$, dictated by the boundary or initial conditions, and the
Kolmogorov scale
\begin{equation}
l_{K}\sim (\nu^3/\epsilon)^{1/4}\ll L_O \ ,
\end{equation}
where $\epsilon$ is the energy dissipation per unit volume \cite{Frisch}.
At the scale $l_K$ the flow becomes smooth (here we make a rough estimate neglecting intermittency), so the Knudsen number obeys
$Kn\sim l_{cor}/l_K$. Using $\epsilon\sim v_c^3/L_O$ where $v_c$ is characteristic value of the fluctuating component of the flow velocity,
one may reexpress $Kn$ in terms of $L_O$ and $v_c$ as
\begin{equation}
Kn\sim (l_{cor}/L_O)^{1/4}(v_c/c)^{3/4} \ .
\end{equation}
Here we used the fact that generally
$F(\lambda)\sim G(\lambda)$, see \cite{FBB,Fouxon:2008tb}. One sees that one can realize $Kn\ll 1$ by dialing the outer scale of turbulence $L_O$,
and the characteristic fluctuating component of the velocity $v_c$.
The possibility of having a globally defined hydrodynamic flow with a uniformly small Knudsen number is specific for non-relativistic
flows, and it normally does not exist for relativistic (compressible) flows, see details in sections \ref{shocks} and  \ref{gravity}.

In this paper we consider the hydrodynamics of relativistic conformal field theories at finite temperature
and its slow motions limit, where it reduces to the incompressible Navier-Stokes equations. The symmetries of the equations and their solutions are analyzed.
We construct exact solutions with finite time singularities of one-dimensional relativistic conformal hydrodynamic motions, and use them to generate multi-dimensional solutions via  special
conformal transformations. These solutions, however, are shown to have no non-trivial slow motions
limit. A simple non-equilibrium steady state in the form of a shock solution is constructed, and its inner structure is analyzed.
We demonstrate that the derivation of the gravitational dual description of conformal hydrodynamics is analogous
to the derivation of hydrodynamics equations from the Boltzmann equation.
The shock solution is shown to correspond to a domain-wall solution in gravity. We show that the solutions to the non-relativistic
incompressible Navier-Stokes equations play a special role in the construction of global solutions to gravity.

The paper is organized as follows. In the section 2 we consider the symmetries of relativistic conformal hydrodynamics and their relation to the symmetries of the incompressible Euler and Navier-Stokes equations in the limit of slow motions.
In section 3 we construct exact solutions of ideal CFT hydrodynamics.
We analyse one-dimensional macroscopic motions, introduce Riemann variables and provide a simple wave solution that becomes singular in a finite time.
We use the solution
in order to generate multi-dimensional solutions via  special
conformal transformations. We show, however, that these solutions have no non-trivial slow motions
limit.
In section 4
we discuss weak solutions of ideal hydrodynamics defined at all times.
We construct
a stationary shock solution as an example of a global weak solution. We describe its inner structure due to viscosity,
its conformal transformation and its non-relativistic limit.
In section 5
we argue that the derivation of the gravitational dual description of conformal hydrodynamics is analogous
to the derivation of hydrodynamics equations from the Boltzmann equation.
We show that the shock solution corresponds to a domain-wall solution in gravity.

\section{Symmetries of CFT hydrodynamics}

The conformal group is the symmetry group of the CFT equations (\ref{cfteq}), and it maps solutions into solutions.
The group includes the Poincare group, a dilatation and four special conformal transformations.
The Poincare group reduces in the limit of slow motions to the Galilean group, including Galilean transformations and space and time translations, which is
a symmetry group of the Navier-Stokes equations.

Consider next the dilatation that acts as:
\begin{equation}
x^{\mu} \rightarrow \lambda x^{\mu},~~~~~u^{\mu}(x^{\alpha})\rightarrow u^{\mu}(\lambda x^{\alpha}),~~~~~T(x^{\alpha}) \rightarrow \lambda^{-1}T(\lambda x^{\alpha}) \ . \label{dilat}
\end{equation}
The dilatation is a symmetry of the CFT equations (\ref{cfteq}), and it remains a symmetry
for ideal hydrodynamics in the limit of non-relativistic macroscopic motions, i.e. a symmetry of the Euler equations.
It acts as
\begin{equation}
t\rightarrow \lambda t,~~~~x^i\rightarrow \lambda x^i,~~~~ v^i\rightarrow v^i \ . \label{t1}
\end{equation}
For viscous hydrodynamics in the limit of non-relativistic macroscopic motions, one has the dimensionfull kinematic viscosity that breaks
this symmetry, and indeed, (\ref{t1}) is not a symmetry of the Navier-Stokes equations.

The Euler equations have a bigger symmetry \cite{Frisch}, which is generated by
(\ref{t1}) together with
\begin{equation}
t\rightarrow \mu t,~~~~x^i\rightarrow x^i,~~~~ v^i\rightarrow \mu^{-1} v^i \ .
\label{t2}
\end{equation}
However, the symmetry (\ref{t2}) does not follow from the symmetry group of relativistic conformal hydrodynamics in
the limit of slow motions, cf. below.
A combination of the two symmetries (\ref{t1}) and (\ref{t2}) with the same transformation parameter $\lambda=\mu=\alpha$
yields
 \begin{equation}
t\rightarrow \alpha^2 t,~~~~x^i\rightarrow \alpha x^i,~~~~ v^i\rightarrow \alpha^{-1} v^i \ ,
\label{t3}
\end{equation}
which is a symmetry of the non-relativistic viscous equations.
In view of the above discussion, we conclude that also this symmetry does not follow from the
symmetry group of relativistic conformal hydrodynamics in
the limit of slow motions.

\subsection{Special Conformal symmetries of relativistic motions}


The special conformal transformations of the space-time coordinates $x^{\mu}$ are given by
\begin{eqnarray}&&
\Phi^{\mu}(x, b)=\frac{x^{\mu}+b^{\mu}x^2}{1+2b\cdot x+b^2x^2} \ ,
\end{eqnarray}
where $b^{\mu}$ is a constant four-vector.
They satisfy the identity
\begin{eqnarray}&&
\eta^{\mu\nu}(\partial_{\mu} \Phi^{\alpha})(\partial_{\nu}\Phi^{\beta})=\frac{\eta^{\alpha\beta}}
{(1+2b\cdot x+b^2x^2)^2} \ . \label{conf1}
\end{eqnarray}

A main distinguishing property of the CFT hydrodynamics is that the special conformal transformations allow to generate new solutions out of a given one.
Consider a stress-energy tensor $T^{\mu\nu}(x)$ of the conformal field theory, i.e. that satisfies
equations (\ref{cfteq}).
Then, special conformal transformations produce a four-parametric family
\begin{eqnarray}&&
T_{\mu\nu}(x, b)=\frac{(\partial_{\mu}\Phi^{\alpha})(\partial_{\nu}\Phi^{\beta})}
{(1+2b\cdot x+b^2x^2)^2}
T_{\alpha\beta}\left[\Phi^{\rho}(x, b)\right] \ . \label{gener}
\end{eqnarray}
where each $T_{\mu\nu}(x, b)$ satisfies equations (\ref{cfteq}).

The vanishing of the
trace $T^{\mu}_{\mu}(x, b)$ follows immediately from Eq.~(\ref{conf1}).
It is also straightforward to check that the equation $\partial_{\nu}T^{\mu\nu}(x, b)=0$ is satisfied, where the derivative
is taken with respect to $x$.
Thus, conformal transformations generate from any solution of (\ref{cfteq})
a four-parameter family of solutions.
Consider now the relativistic conformal hydrodynamics, where the stress-energy tensor
is defined by the four fields: $u^{\mu}$ and $T$.
The analysis above implies that special conformal transformations of any solution $u^{\mu}$, $T$ of the
conformal hydrodynamic equations generate a four-parameter family of solutions $T(x, b)$, $u_{\mu}(x, b)$, where
\begin{eqnarray}&&
u_{\mu}(x, b)=(1+2b\cdot x+b^2x^2)(\partial_{\mu}\Phi^{\alpha})u_{\alpha}\left[\Phi(x, b)\right],\label{velocity}\\&& T(x, b)=\frac{T\left[\Phi(x, b)\right]}{1+2b\cdot x+b^2x^2} \ . \label{temperature}
\end{eqnarray}
One can verify that $u_{\mu}(x, b)$ has the correct normalization $u_{\mu}(x, b)u_{\nu}(x, b)\eta^{\mu\nu}=-1$ and the above transformations of $u$ and $T$ lead to
transformation (\ref{gener}) both for the ideal hydrodynamics $T^{\mu}$ given by Eq.~(\ref{ideal00}) and
for viscous hydrodynamics $T^{\mu\nu}$ given by Eq.~(\ref{visc00}).
Note that the velocity transformation (\ref{velocity}) can be rewritten
as
\begin{eqnarray}&&
u^{\mu}(x, b)=u^{\mu}\left[\Phi(x, b)\right]+2u^{\alpha}\left[\Phi(x, b)\right]
\nonumber\\&&
\times\frac{b_{\alpha}x^{\mu}-b^{\mu}x_{\alpha}
+2b_{\alpha}x^{\mu}b\cdot x-b^2x_{\alpha}x^{\mu}-x^2b_{\alpha}b^{\mu}}
{1+2b\cdot x+b^2x^2} \ .\label{velocity1}
\end{eqnarray}

\subsection{Special conformal symmetries of non-relativistic motions}

In order to analyze the non-relativistic limit of special conformal transformations
of the hydrodynamics it is convenient to consider
the special conformal transformation of $\bm v$ defined by $u_{\alpha}=(-\gamma, \gamma \bm v/c)$. Using
$v_i=-cu_i/u_0$,  one finds that Eq.~(\ref{velocity}) implies
the transformation of
$\bm v$
\begin{eqnarray}&&
\!\!\!\!\!\!v_i(x, b)\!=\!\frac{(1+2b\cdot x+b^2x^2)v_i\left[\Phi(x, b)\right]+2v_j\left[\Phi(x, b)\right]\left[b^{j}x_{i}-b_{i}x^{j}
+2b^{j}x_{i}b\cdot x-b^2x^{j}x_{i}-x^2b^{j}b_{i}\right]}{1+2b\cdot x+b^2x^2+
2\left[2b^0x_0 b\cdot x+b^2x^2_0+x^2b^2_0\right]-2v_j\left[\Phi(x, b)\right]\left[b^{j}x_{0}-b_{0}x^{j}
+2b^{j}x_{0}b\cdot x-b^2x^{j}x_{0}-x^2b^{j}b_{0}\right]/c}\nonumber\\&&
-\frac{2c\left[b^{0}x_{i}-b_{i}x^{0}
+2b^{0}x_{i}b\cdot x-b^2x^{0}x_{i}-x^2b^{0}b_{i}\right]}{1+2b\cdot x+b^2x^2+
2\left[2b^0x_0 b\cdot x+b^2x^2_0+x^2b^2_0\right]-2v_j\left[\Phi(x, b)\right]\left[b^{j}x_{0}-b_{0}x^{j}
+2b^{j}x_{0}b\cdot x-b^2x^{j}x_{0}-x^2b^{j}b_{0}\right]/c} \label{form} \ .
\end{eqnarray}
Consider solutions of relativistic conformal hydrodynamics, where $\bm v$ remains finite in the limit $c\to\infty$, and thus obeys the non-relativistic incompressible Euler (Navier-Stokes)
equations. Then, if we employ in the above transformation such parameters $b$ that $\bm v(x, b)$ is also finite at $c\to\infty$, we find a symmetry of the Euler equation. First it
is necessary that the argument $\Phi(x, b)$ is finite in the limit $c\to\infty$.
This implies $b^i=a^i/c^2$ and $b^0=a/c$ where $a^i$ and $a$ have either finite or zero limit at $c\to\infty$. Thus,
\begin{equation}
t \rightarrow \frac{t}{1-a t},~~~~~~x^i \rightarrow \frac{x^i - a^i t^2}{(1 - at)^2} \ .
\label{spnonr}
\end{equation}

The expression (\ref{form}) gives the transformation of $\bm v$ in the non-relativistic limit
\begin{eqnarray}&&
\bm v(\bm x, t, a, \bm a)=\bm v\left[\frac{t}{1-at}, \frac{\bm x-\bm a t^2}{(1-at)^2}\right]
+2\frac{\bm a t-a\bm x}{1-at} \ .
\end{eqnarray}
If we take $a=0$ (or equivalently assume that $a$ scales as some negative power of $c$), while $a_i$ is finite at $c\to\infty$, then we get the following symmetry of the Euler equation
\begin{eqnarray}&&
\bm v(\bm x, t, \bm a)=\bm v\left[\bm x-\bm a t^2, t\right]+2\bm a t \ .
\end{eqnarray}
This symmetry, discussed also in \cite{Bhattacharyya:2008kq},
corresponds to adding a uniform gravitational field to the Euler equation. Indeed, if $\bm v(\bm x, t)$ satisfies
the incompressible Euler equation then $\bm v(\bm x, t, \bm a)$ satisfies $\partial_t\bm v+(\bm v\cdot\nabla)\bm v=-\nabla p+2\bm a$. However $2\bm a=\nabla (2\bm a\cdot\bm r)$ can be reabsorbed in the
pressure, so that $\bm v(\bm x, t, \bm a)$ is also a solution of the incompressible Euler equation. Thus,this symmetry corresponds to
a combination of a passage to a uniformly accelerating frame and the possibility of reabsorbing a (uniform) gravitational field into the
pressure (known in fluid mechanics see e.g. pp. $16-17$ in \cite{Pope}). The above consideration also works for the Navier-Stokes equation.

If we choose $a^i=0$ (equivalent to assuming that $a_i$ decreases as a negative power of $c$), while $a$ remaining finite at $c\to\infty$ we get
\begin{eqnarray}&&
\bm v(\bm x, t, a)=\bm v\left[\frac{t}{1-at}, \frac{\bm x}{(1-at)^2}\right]-
\frac{2a\bm x}{1-at}. \label{temporal}
\end{eqnarray}
This transformation describes an addition of a total expansion (from the center at $\bm x=0$)
or a total contraction (to the center at $\bm x=0$) to the given flow. However, this transformation
does not preserve the condition of finite velocity at large distances and as a result is
not a symmetry of the incompressible Euler equation. In particular, out of an
incompressible velocity it produces a field with $\nabla\cdot \bm v(\bm x, t, a)=6a/(1-at)$.

Although the transformation (\ref{temporal}) breaks the finiteness of velocity at infinity, and does not describe a symmetry of the incompressible Euler equation,
it is a symmetry of the more
general equation (\ref{Eul1}). Note, also that this transformation is different from the special conformal transformation that completes the Schr\"{o}dinger
symmetry group. The latter acts as
 \begin{equation}
t \rightarrow \frac{t}{1-a t},~~~~~~x^i \rightarrow \frac{x^i}{1 - at} \ .
\label{spnonr}
\end{equation}


\section{Exact solutions of ideal CFT hydrodynamics}

\subsection{One-dimensional motions and finite time singularities in ideal CFT hydrodynamics}

Equations (\ref{cfteq}) can be written as \cite{Son}
\begin{eqnarray}&&
{\cal D} \xi=-\frac{1}{3}\partial_{\nu} u^{\nu},\ \ {\cal D}u^{\mu}=-\partial^{\mu}\xi+\frac{u^{\mu}\partial_{\nu} u^{\nu}}
{3}, \label{a1}
\end{eqnarray}
where $\xi\equiv \ln T$ and ${\cal D}=u^{\alpha}\partial_{\alpha}$.
The first equation follows from the second by multiplication
with $u_{\mu}$ and the use of $u^{\mu}u_{\mu}=-1$.

Let us consider one-dimensional hydrodynamic motions, that is motions that depend only on one
spatial coordinate, say $x$, and have $u^y=u^z=0$. It should be stressed that the microscopic dynamics of the system
is still $(3+1)-$dimensional, and it is only the macroscopic motion that is one-dimensional.
For example, one could think of pushing the CFT fluid through a pipe where $x$ is
the coordinate along the pipe.
It is convenient to parameterize the velocity by $u^0=\cosh \phi$ and $u^x=\sinh \phi$.
Substituting $u^{\mu}=(\cosh \phi, \sinh \phi, 0, 0)$, we find that the
$\mu=0$ and $\mu=1$ components of the second of Eqs.~(\ref{a1}) give
\begin{eqnarray}&&
\frac{\partial \xi}{\partial t}=\frac{\sinh 2\phi}{3}\frac{\partial \phi}{\partial t}
+c\frac{\cosh 2\phi-2}{3}\frac{\partial \phi}{\partial x}\ , \label{syst1} \\&&
-\frac{\partial \xi}{\partial x}=\frac{\cosh 2\phi+2}{3c}\frac{\partial \phi}{\partial t}
+\frac{\sinh 2\phi}{3}\frac{\partial \phi}{\partial x}, \label{syst2}
\end{eqnarray}
respectively. This  system has a more symmetric form when written as an evolution equation
for $\xi$ and $\phi$,
\begin{eqnarray}&&
\frac{\partial \xi}{\partial t}=-\frac{c\sinh(2\phi)}{\left[\cosh(2\phi)+2\right]}\frac{\partial \xi}{\partial x}-\frac{c}{\left[\cosh(2\phi)+2\right]}\frac{\partial \phi}{\partial x},\label{syst5}\\&&
\frac{\partial \phi}{\partial t}=-\frac{c\sinh(2\phi)}{\left[\cosh(2\phi)+2\right]}\frac{\partial \phi}{\partial x}-\frac{3c}{\left[\cosh(2\phi)+2\right]}\frac{\partial \xi}{\partial x}.\label{syst6}
\end{eqnarray}
It is interesting to notice the particular form that the above equations take, when considered
up to quadratic order in $\phi$. Introducing $v'/c=2\phi/3$ and $\rho=T^2$ we find
\begin{eqnarray}&&
\frac{\partial \rho}{\partial t}+\frac{\partial (\rho v')}{\partial x}=0,\\&&
\rho\left(\frac{\partial v'}{\partial t}+v'\frac{\partial v'}{\partial x}\right)=-\frac{\partial p}{\partial x},\
\ p\equiv\frac{\rho c^2}{3} \ .
\end{eqnarray}
These equations describe an isothermal one-dimensional flow of gas with temperature $c^2/3$.
They also describe one-dimensional isentropic, i.e constant entropy density, motion of a barotropic gas with
velocity $v'$ and pressure $p=\rho c^2/3$. In general, a barotropic gas with isentropic motion obeys $p=A\rho^{\gamma}$
where $A$ is a constant and $\gamma\geq 1$ is the polytropic exponent of the gas, which is equal to the ratio of
the heat capacities at constant pressure and volume, respectively. Here $\gamma=1$, which
corresponds
to gas in the limit of a large number of internal degrees of freedom of the molecules.
It is notable that a motion of such a gas admits a gravity dual interpretation, as we will discuss in the last section.

The system of equations (\ref{syst1})-(\ref{syst2}) is a $2\times 2$ system of quasi-linear PDE. It
can be recast using Riemann variables, which are conserved along
characteristic directions in space \cite{Whitham}. Here the result of the procedure can be readily seen directly from the equations. Multiplying the second of Eqs.~(\ref{syst5})-(\ref{syst6}) with $1/\sqrt{3}$ and adding and subtracting it from the first equation we get
\begin{eqnarray}&&
\frac{\partial \xi}{\partial t}+\frac{c\left[\sinh 2\phi\pm \sqrt{3}\right]}{\cosh 2\phi+2}\frac{\partial \xi}{\partial x}\pm \frac{1}{\sqrt{3}}
\Biggl[\frac{\partial \phi}{\partial t}+\frac{c\left[\sinh 2\phi\pm \sqrt{3}\right]}{\cosh 2\phi+2}\frac{\partial \phi}{\partial x}\Biggr]=0.
\end{eqnarray}
These equations relate derivatives of $\phi$ and $\xi$ along the two families $x_{\pm}(t)$
of characteristic directions, defined by
\begin{eqnarray}&&
\frac{dx_{\pm}}{dt}=\frac{c\left[\sinh 2\phi\pm \sqrt{3}\right]}{\cosh 2\phi+2} \ .
\end{eqnarray}
Thus, two families of characteristics emanate from each point of the fluid. They describe non-linear propagation of sound to the right ($x_+$)and to the left ($x_{-}$). In the limit of linear sound (small $\phi$), the equations are simply
${\dot x}_{\pm}=\pm c/\sqrt{3}$ where $c/\sqrt{3}$ is the speed of sound.
Correspondingly, the Riemann variables have a remarkably simple form
\begin{eqnarray}&&
r_{\pm}=\xi\pm\frac{\phi}{\sqrt{3}},
\end{eqnarray}
and the equations governing one-dimensional hydrodynamic motions of the CFT fluid can be written
in the form
\begin{eqnarray}&&
\left[\frac{\partial}{\partial t}+c\frac{\sinh [\sqrt{3}(r_+ -r_-)]+\sqrt{3}}{\cosh [\sqrt{3}(r_+ -r_-)]+2}\frac{\partial}{\partial x}\right]r_+=0, \label{Riemann1}\\&&
\left[\frac{\partial}{\partial t}+c\frac{\sinh [\sqrt{3}(r_+ -r_-)]-\sqrt{3}}{\cosh [\sqrt{3}(r_+ -r_-)]+2}
\frac{\partial}{\partial x}\right]r_-=0 \ . \label{Riemann2}
\end{eqnarray}
The above system admits special solutions, called simple waves, where one of Riemann variables is a
constant. For example, for $r_-=const$, we find a closed equation on $r_+$ of the form
\begin{equation}
\partial_t r_++f(r_+)\partial_x r_+=0 \ .
\end{equation}
 Introducing $w=f(r_+)$ we find that $w$ obeys the
Hopf equation
\begin{eqnarray}&&
\frac{\partial w}{\partial t}+w\frac{\partial w}{\partial x}=0 \ . \label{Hopf}
\end{eqnarray}
The equation describes formation of discontinuities (shocks) in $w$ in a finite time. The shocks
form when the spatial derivative of the initial condition on $w$ is negative somewhere. Indeed, introducing
$\sigma=\partial_x w$ we find that $\sigma$ obeys
\begin{eqnarray}&&
\frac{d}{dt}\sigma[x(t), t]=-\sigma^2[x(t), t],\ \ \frac{dx}{dt}=w[x(t), t]. \label{Hopf1}
\end{eqnarray}
This equation predicts finite time blowup of $\sigma$, that occurs at $x(t)$ emanating  from
the initial point where $\sigma$ has the (negative) minimum. It is easy to see that the blowup
corresponds to the formation of a discontinuity in $w$. One can use the implicit form of the solution
of Eq.~(\ref{Hopf}),
\begin{eqnarray}&&
w(x, t)=w_0\left[x-w(x, t)t\right],\ \ w_0\equiv w(x, t=0).
\end{eqnarray}
Qualitatively, Eq.~(\ref{Hopf}) describes
velocity field of free particles moving on a line where the discontinuity forms when
faster particles catch up with the slower particles in front of them.

The formal reason for the formation of discontinuities in a simple wave is the intersection
of characteristics. The field $w$ in Eq.~(\ref{Hopf}) is conserved along the characteristic lines
$x(t)$ defined in Eq.~(\ref{Hopf1}). In general, nothing prevents the lines from intersecting
in a finite time $T$. At $t>T$ the solution obtained by imposing the field conservation along
the characteristics would become multi-valued, which is signaled by the divergence of the spatial
derivative of $w$ at the intersection point. The same property of the finite-time formation of infinite
derivatives of the fields holds for the solutions of  the system (\ref{Riemann1})-(\ref{Riemann2})
and thus, also of the original system (\ref{syst1})-(\ref{syst2}).
Indeed, the equations imply that $r_{\pm}$ are conserved along the corresponding characteristics in $(x, t)$
plane. Generically, these characteristics intersect in a finite time, and at the intersection
time a field derivative diverges. Thus, we conclude that ideal hydrodynamics equations in one dimension
lead to a finite time singularity of the derivatives.

\subsection{Special conformal symmetry generation of higher-dimensional solutions}

As shown
in the previous section, one may solve rather generally the one-dimensional dynamics of the fluid in a CFT hydrodynamics.
We can generate from the one-dimensional solutions, higher dimensional ones by applying
special conformal
transformation as in Eqs.~(\ref{velocity})-(\ref{temperature}).
With the help of Eq.~(\ref{velocity1}), one finds that a solution of one-dimensional hydrodynamics leads
to a family of solutions:
\begin{eqnarray}&&
u^0(x, b)=\cosh \phi\left[\Phi(x, b)\right]\left[1+2\frac{(\bm b x^0-\bm x b^0)^2}
{1+2b\cdot x+b^2x^2}\right]
\nonumber\\&&
+\frac{2\sinh \phi\left[\Phi(x, b)\right]}{1+2b\cdot x+b^2x^2}\Biggl[b_1x^0-b^0x_1
+2b_1x^0b\cdot x-b^2x_1x^0 \nonumber\\&&-x^2b_1b^0\Biggr],\ \
u^1(x, b)=\sinh \phi\left[\Phi(x, b)\right]\Biggl[1
\nonumber\\&&
+2\frac{2b^1x^1 b\cdot x-b^2(x^1)^2-x^2(b^1)^2}
{1+2b\cdot x+b^2x^2}\Biggr]+2\cosh \phi\left[\Phi(x, b)\right]\nonumber\\&&\times\frac{b_0x^1-b^1x_0
+2b_0x^1b\cdot x-b^2x_0x^1-x^2b_0b^1}
{1+2b\cdot x+b^2x^2},\label{generalstr0}\\&&
u^2(x, b)=\frac{2}{1+2b\cdot x+b^2x^2}\Biggl[
\cosh \phi\left[\Phi(x, b)\right]\Biggl(b_0x_2-b_2x_0\nonumber\\&&
+2b_0x_2b\cdot x-b^2x_0x_2-x^2b_0b_2\Biggr)+\sinh \phi\left[\Phi(x, b)\right]\nonumber
\\&&\times\left(b_1x_2-b_2x_1
+2b_1x_2b\cdot x-b^2x_1x_2-x^2b_1b_2\right)\Biggr]\nonumber
\\&&
\xi(x, b)=\xi\left[\Phi(x, b)\right]+\ln\left[1+2b\cdot x+b^2x^2\right],
\label{generalstr}
\end{eqnarray}
where $b^{\mu}=(b^0, \bm b)$ and $x^{\mu}=(x^0, \bm x)$, the expression for $u^3$ is obtained
from the expression for $u^2$ by interchange of indices, and $\phi$ and $\xi$ are obtained
from the one-dimensional hydrodynamics.

\subsection{The non-relativistic limit}

The relativistic one-dimensional solution does not have a non-trivial slow motions limit since the
incompressibility condition implies that the velocity is independent of the coordinate $x$. This is
in accord with the presence of a factor of $c$ in front of the space derivative in Eqs.~(\ref{Riemann1})-(\ref{Riemann2}).
We have seen that we can generate from the relativistic one-dimensional solution, higher dimensional ones by applying special
conformal transformations.
It is still natural to ask whether these higher-dimensional solutions have a non-trivial slow motions
limit, i.e. whether they also solutions of the incompressible Navier-Stokes equations.
We will show in the following that this is not the case, and also the higher-dimensional solutions do not have
a non-trivial slow motions limit.
Of course, this is not really a surprise, since as we noted above, the one-dimensional solution with which we started in order to generate
the higher-dimensional ones, has a trivial limit.

For simplicity we set the time coordinate to zero.
One can write the general structure of the velocity $v^i$ using equations (\ref{generalstr}) as
\begin{equation}
v_i(r) = (1 +2b\cdot \bm x+b^2 {\bm x}^2)\phi(\Phi)\nabla_i \Phi  \ ,
\end{equation}
where
\begin{equation}
\Phi = \frac{x +b^{x}{\bm x}^2}{1+2b\cdot \bm x+b^2 {\bm x}^2} \ .
\end{equation}
The incompressibility condition implies then that
\begin{equation}
\frac{\phi'}{\phi}\left( \Phi \right) = - \frac{\nabla_i \left[(1+2b\cdot \bm x+b^2{\bm x}^2)\nabla_i \Phi \right)}{(1+2b\cdot\bm x +b^2{\bm x}^2)(\nabla \Phi)^2} \ .
\label{consis}
\end{equation}
Since the LHS of (\ref{consis}) is a function of $\Phi$, we need to require that the RHS is also a function of
$\Phi$ in order to be able to solve of $\phi$. However, a straightforward calculation shows that the RHS of (\ref{consis}) is
not a function of $\Phi$, for $\Phi$ finite in the slow motions limit.
Thus, the incompressibility equation cannot be solved, and there is no non-trivial slow motion limit to the
higher-dimensional solution of the relativistic conformal hydrodynamics.

\section{Shock solutions}
\label{shocks}

\subsection{Shocks - global solutions in ideal hydrodynamics}

We have seen that the relativistic ideal conformal hydrodynamics evolution in one dimension produces discontinuities in a finite time. The mechanism for
the generation of the singularity is the intersection of characteristics, and it holds in higher-dimensional motions as well. A
simple concrete example is the multi-dimensional generalization of the Hopf equation (\ref{Hopf}) provided by $\partial_t  \bm w+\bm w\nabla\bm w=0$. The matrix $\sigma_{ij}=\partial_j w_i$ obeys the equation $d\sigma/dt=-\sigma^2$ , that predicts a finite time blow up of $\sigma$. The finite-time generation of discontinuities is a general property of ideal hydrodynamics.

A natural question that arises is how should we define the hydrodynamics evolution beyond the singularity time?
One way is to include the viscous contribution to the stress tensor,  and consider Eqs.~(\ref{cfteq}) and (\ref{visc00}) instead of Eqs.~(\ref{cfteq}) and (\ref{ideal00}). According to the ideal hydrodynamics equations,
the viscous contribution grows in time and becomes infinite at the time of formation of the discontinuity
[cf. Eq.~(\ref{str})]. Thus, the viscous contribution is important near the singularity. It is conjectured (cf. \cite{Landau,Frisch}), that this contribution regularizes the  singularity,
 makes all the fields finite and provides a well-defined evolution within
the frame of differential equations.

A different approach for dealing with the singularity is to continue the solutions of the ideal
hydrodynamics beyond the singularity time,  by passing from classical (differentiable) to weak solutions of the equations, where
weak solutions are solutions with discontinuities. This formulation is very useful when considering
high Reynolds number flows, for which the viscosity is effectively small \cite{Landau}. In such
flows the viscosity is important only near the points, where a
discontinuity would form in the absence of viscosity. The effect of viscosity is to smooth the discontinuity, while  in the other regimes of the
flow the viscosity effect is small.

An appropriate framework is to apply an
ideal hydrodynamics evolution everywhere, except at the hypersurfaces where a discontinuity forms. At the discontinuity one requires that the normal components of
 fluxes of locally conserved energy and momentum densities remain continuous. This requirement
 is necessary in order for the equations to
describe the local conservation laws, as the hydrodynamics should do. The corresponding constraints at the
discontinuity are called the Rankine-Hugoniot conditions.

The main advantage of the obtained weak solutions - the shocks - is that they provide within the framework of ideal hydrodynamics,
global and valid solutions at all times. Let us construct a simple example
of such a solution.
The simplest solution of the system of Eqs.~(\ref{cfteq}) and (\ref{ideal00}) is a boosted equilibrium state, where
$T$ and $u^{\mu}$ are constants. We can saw two equilibrium solutions along
a plane, and  form a simple but non-trivial solution - a shock wave.
$T$ and $u^{\mu}$ are piecewise constant and they have a single discontinuity along a plane
$x=x(t)$. In the frame that moves with the shock where $x(t)\equiv 0$ (the considered shock solutions move at the constant speed), the fluxes of the conserved quantities must be continuous
through the shock. This leads to the Rankine-Hugoniot conditions
\begin{eqnarray}&&
[T^4\left(\eta^{\mu x}+4u^{\mu}u^{x}\right)]=0 \ ,
\label{boundary}
\end{eqnarray}
where the brackets stand for the value of the jump in the function through the shock.
In contrast, the entropy flux does not have to be continuous - rather the entropy grows in the shock.

Using $u^{\mu}=(\gamma, \gamma \bm v)$ (here we set $c=1$), we get from Eq.~(\ref{boundary})
\begin{eqnarray}&&
\left[\frac{T^4v_x}{1-v^2}\right]=0,\ \ \left[T^4+\frac{4T^4v_x^2}{1-v^2}\right]=0  \ ,
\end{eqnarray}
and
\begin{eqnarray}&&
\left[\frac{T^4v_yv_x}{1-v^2}\right]=0,\ \ \left[\frac{T^4v_zv_x}{1-v^2}\right]=0.
\end{eqnarray}
If $v_x$ is non-zero the equations imply that the velocity components parallel to the
shock plane are continuous across the shock. For convenience, we can set these components to zero, and we are left with
\begin{eqnarray}&&
\left[\frac{T^4v_x}{1-v_x^2}\right]=0,\ \ \left[T^4+\frac{4T^4v_x^2}{1-v_x^2}\right]=0 \ .
\end{eqnarray}
Note, that if we use the general conditions for the shock \cite{Landau} with $p=aT^4$ and heat
function $w=4aT^4$ (where $a=const$), then we get the same conditions.

If we denote by $T_1, T_2$ and $v_1,v_2$ the values of the temperature and the velocity component $v_x$ at two sides
of the shock, then the conditions read
\begin{eqnarray}&&
\frac{T_2^4 v_2}{1-v_2^2}=\frac{T_1^4 v_1}{1-v_1^2},\ \ T_2^4+\frac{4T_2^4 v_2^2}{1-v_2^2}=T_1^4+\frac{4T_1^4 v_1^2}{1-v_1^2}. \label{RH}
\end{eqnarray}
These equations define a two-parameter family of shock solutions of hydrodynamics. The two parameters
correspond to the values of the fluxes of the energy and the $x-$component of the momentum through the shock. Evidently, $T_1=T_2$
and $v_1=v_2$ solve the above equations. In order to analyze the conditions for a non-trivial solution,
 we assume that the fields $T_2$, $v_2$ at $x<0$ are given (one can assume $v_2>0$ as the solution with $v_2<0$ can be obtained by the sign reversal).
Equating the expressions for $T_2^4/T_1^4$, we find a quadratic equation for $v_1$ that has two solutions, $v_1=\frac{1}{3v_2}$ and the trivial solution $v_1=v_2$. Thus, a non-trivial solution with
$v_1= \frac{1}{3v_2} \leq 1$ exists only for $v_2$ above the threshold value of $1/3$ (recall that the speed of sound is $1/\sqrt{3}$).
In the frame
for which the fluid is at rest at $-\infty$, the shock always moves at a speed larger than $1/3$.
Note that
the difference $v_1-v_2$ can be arbitrarily
small.

Fixing the temperature from Eq.~(\ref{RH}), we find that non-trivial
shock solutions at $v_2>1/3$ are described by the relations
\begin{eqnarray}&&
v_1=\frac{1}{3v_2},\ \ T_1^4=T_2^4\frac{9v_2^2-1}{3(1-v_2^2)} \ . \label{idshock}
\end{eqnarray}
It is remarkable that the Rankine-Hugoniot conditions for the CFT have such explicit solutions, cf. the usual situation \cite{Landau}.
Note that $v_1>v_2$ for $1/3<v_2<1/\sqrt{3}$ while $v_1<v_2$ for $v_2>1/\sqrt{3}$ where $1/\sqrt{3}$ is the speed of sound. For $v_2$ close
to the speed of sound we have $v_1\approx v_2$ while at the exact equality $v_2=1/\sqrt{3}$ the solution trivializes.
Finally $v_1$, like $v_2$,
belongs to the range $1/3<v_1<1$ as should be due to the symmetry in the indices interchange.

The above shock solutions are stationary non-equilibrium states of the fluid with finite dissipation in the shock. The stationarity is sustained by the fact that the solution is non-vanishing at infinity, and the presence of finite fluxes in the system.

\subsection{Special conformal transformation of shocks}

Shocks are quite simple solutions of ideal hydrodynamics: they are time-independent, one-dimensional and  piecewise constant. As a result they allow a clear insight into what kind of a change the special conformal transformations produce in the solution. Using Eqs.~(\ref{form}) and (\ref{temperature}) we find
that shocks produce the following weak solution of the hydrodynamics
\begin{eqnarray}&&
\!\!\!\!\!\!v_i(x, b)\!=\!\frac{(1+2b\cdot x+b^2x^2)v_{[k]}\delta_{i1}+2v_{[k]}\left[b^{1}x_{i}-b_{i}x^{1}
+2b^{1}x_{i}b\cdot x-b^2x^{1}x_{i}-x^2b^{1}b_{i}\right]}{1+2b\cdot x+b^2x^2+
2\left[2b^0x_0 b\cdot x+b^2x^2_0+x^2b^2_0\right]-2v_{[k]}\left[b^{1}x_{0}-b_{0}x^{1}
+2b^{1}x_{0}b\cdot x-b^2x^{1}x_{0}-x^2b^{1}b_{0}\right]/c}\nonumber\\&&
-\frac{2c\left[b^{0}x_{i}-b_{i}x^{0}
+2b^{0}x_{i}b\cdot x-b^2x^{0}x_{i}-x^2b^{0}b_{i}\right]}{1+2b\cdot x+b^2x^2+
2\left[2b^0x_0 b\cdot x+b^2x^2_0+x^2b^2_0\right]-2v_{[k]}\left[b^{1}x_{0}-b_{0}x^{1}
+2b^{1}x_{0}b\cdot x-b^2x^{1}x_{0}-x^2b^{1}b_{0}\right]/c}, \\&&
T(x, b)=\frac{T_{[k]}}{1+2b\cdot x+b^2x^2}.
\end{eqnarray}
where $[k]=1, 2$ and $v_{[k]}$, $T_{[k]}$ are some constants obeying Eqs.~(\ref{RH}). The equation of discontinuity hyperplane (where one should switch $k=1$ to $k=2$ in the above formula) is given by $x^{1}+b^{1}x\cdot x=0$.
One observes that the special conformal transformations allow to generate quite complicated motions out of
even relatively simple motions.

\subsection{Shocks in viscous CFT hydrodynamics}

The ideal hydrodynamics shock solution constructed above is clearly not uniformly valid in space. It contains a discontinuity region, where
derivatives are formally infinite indicating that one must account for the viscous, and maybe also higher order derivatives contributions
to $T^{\mu\nu}$. The ideal hydrodynamics shock describes correctly the exact solution  far from the region of fast
variations of the fields, and it arises when the formal limit of zero viscosity is taken in the exact solution.
Here we provide the exact solution of the viscous hydrodynamics, and demonstrate how the one-dimensional shock considered in the first subsection arises in the limit.

The stationary one-dimensional solutions of the hydrodynamics equations satisfy
\begin{equation}
\partial_{\nu}T^{\mu\nu}= \partial_0 T^{\mu 0}+\partial_xT^{\mu x}= 0,~~~~~ \partial_0T^{\mu 0}=0 \ .
\end{equation}
These equations imply that
\begin{eqnarray}&&
T^{0x}=C_1,\ \ T^{xx}=C_2 \ , \label{basic110}
\end{eqnarray}
where $C_1$ and $C_2$ are the constant  fluxes of energy and momentum, respectively.

As already
mentioned above, the states in question are non-equilibrium steady states sustained by the presence of finite fluxes in the system. In an ideal hydrodynamics, the non-trivial solutions of the above equations with different values of velocity and temperature at $x=-\infty$ and $x=+\infty$, are obtained by introducing a discontinuity at $x=0$. In contrast, in the viscous hydrodynamics the different asymptotic values at $x=-\infty$ and $x=+\infty$ are connected smoothly and the discontinuity at $x=0$ changes to a smooth transition from the values of the fields at $x=-\infty$ to those at $x=+\infty$.
 Note,
that
at large $|x|$ the derivatives are small and  ideal hydrodynamics is a good approximation. The regions that are
influenced by the viscosity are at the core of the shock.

In order to take into account viscosity,  we consider $\sigma_{\mu\nu}$ in Eqs.~(\ref{visc00})-(\ref{str}). One notes that the time-derivatives appearing in $\sigma_{\mu\nu}$ can be substituted by their
ideal hydrodynamics expressions. Indeed, within hydrodynamics the time derivatives of fields are given
by a series in the spatial gradients of the fields. The second derivative terms in $\sigma_{\mu\nu}$
are higher order terms in the expansion compared to the considered order involving the first derivatives \cite{Weinberg}. The use of Eqs.~(\ref{a1}) allows to rewrite Eq.~(\ref{str}) as \cite{Fouxon:2008tb}
\begin{eqnarray}&&
\sigma_{\mu\nu}=\partial_{\mu}u_{\nu}+\partial_{\nu}u_{\mu}-u_{\nu}\partial_{\mu}\xi-u_{\mu}\partial_{\nu}\xi -2\eta_{\mu\nu}\partial_{\alpha}u^{\alpha}/3,\nonumber
\end{eqnarray}
which gives
\begin{eqnarray}&&
\sigma_{0x}=\cosh \phi\partial_t\phi-\sinh \phi\partial_x \phi+\cosh\phi\partial_x\xi-\sinh \phi\partial_t\xi, \\&&
\sigma_{xx}=4\cosh\phi\partial_x\phi/3-2\sinh\phi\partial_x\xi-2\sinh\phi\partial_t\phi/3.
\end{eqnarray}

The above expressions contain time-derivatives that make the equation $\partial_{\nu}T^{\mu\nu}=0$ a second order equation in the time derivative.
In order to obtain a well-defined evolution equation, which is first-order in time, one has to use the ideal hydrodynamics equation and express the time derivatives via space derivatives. Using
Eqs.~(\ref{syst5})-(\ref{syst6}) one finds
\begin{eqnarray}&&
\sigma_{0x}=-\frac{2w\sinh 2\phi}{2+\cosh 2\phi},\ \
\sigma_{xx}=\frac{2w[1+\cosh 2\phi]}{2+\cosh 2\phi} \ ,\\&& w=\cosh\phi\frac{\partial \phi}{\partial x}-\sinh\phi\frac{\partial \xi}{\partial x}=T\frac{\partial}{\partial x}\left(\frac{\sinh \phi}{T}\right).
\end{eqnarray}
Note, that both components involve the same combination of derivatives $w$. This combination contains besides the derivative
of the velocity also the derivative of the temperature. In non-relativistic hydrodynamics such terms arise
from the heat conduction. Here their appearance is in accord with the fact that the fluid viscosity
allows to capture in a unified way both the effects of viscosity and the heat conduction.
With viscosity the components of the stress-energy tensor read
\begin{eqnarray}&&
T^{0x}=2T^4\sinh 2\phi-\frac{2F(\lambda)T^4\sinh 2\phi}{2+\cosh 2\phi}\frac{\partial}{\partial x}\left(\frac{\sinh \phi}{T}\right),\\&& T^{xx}=T^4\left[2\cosh2\phi-1\right]-\frac{2F(\lambda)T^4[1+\cosh 2\phi]}{2+\cosh 2\phi}\frac{\partial}{\partial x}\left(\frac{\sinh \phi}{T}\right).
\end{eqnarray}

Thus,
stationary one-dimensional solutions of the CFT hydrodynamics obey (cf. Eqs.~(\ref{basic110}) and note that partial derivatives becomes ordinary ones):
\begin{eqnarray}&&
2T^4\sinh 2\phi-\frac{2F(\lambda)T^4\sinh 2\phi}{2+\cosh 2\phi}\frac{d}{dx}\left(\frac{\sinh \phi}{T}\right)=C_1, \label{visc5}\\&& T^4\left[2\cosh2\phi-1\right]-\frac{2F(\lambda)T^4[1+\cosh 2\phi]}{2+\cosh 2\phi}\frac{d}{dx}\left(\frac{\sinh \phi}{T}\right)=C_2 \ .\label{visc6}
\end{eqnarray}
These equations pass to the ideal hydrodynamics equations in the zero viscosity limit
$F(\lambda)\to 0$.
Since far from $x=0$ the viscous contribution in these equations can be neglected due to the smallness of the derivatives, we find that the asymptotic values of the functions at $+\infty$ and $-\infty$ satisfy
\begin{eqnarray}&&
T_{+\infty}^4\sinh 2\phi_{+\infty}=T_{-\infty}^4\sinh 2\phi_{-\infty}=C_1/2 \ ,\ \
T_{+\infty}^4\left[2\cosh2\phi_{+\infty}-1\right]=T_{-\infty}^4\left[2\cosh2\phi_{-\infty}-1\right]=C_2 \ .\label{eq111}
\end{eqnarray}
These conditions with $v=\tanh\phi$ correspond to the Rankine-Hugoniot conditions (\ref{RH}). In this
way the shock characteristics $T_i$ and $v_i$ in Eq.~(\ref{RH}) describe the asymptotic values of the fields far from the region, where viscosity is important. In order
to construct the values of the fields in the viscous region, which corresponds
to a discontinuity in the shock description, one has to solve the system of Eqs.~(\ref{visc5})-(\ref{visc6}). Multiplying the first equation by
$1+\cosh 2\phi$, the second by $\sinh 2\phi$, and subtracting the equations we find
\begin{eqnarray}&&
T^4=\frac{C_1\coth \phi-C_2}{3} \ . \label{eq101}
\end{eqnarray}
Plugging the above relation into Eq.~(\ref{visc5}) we find that $\phi$ satisfies
\begin{eqnarray}&&
\frac{2\sinh 2\phi\left[C_1\coth \phi-C_2\right]}{3}\left[1-\frac{3^{1/4}F(\lambda)}{2+\cosh 2\phi}\frac{d}{dx}\left(\frac{\sinh \phi}{\left[C_1\coth \phi-C_2\right]^{1/4}}\right)\right]=C_1 \ . \label{eq100}
\end{eqnarray}
We seek a solution to this equation that satisfies Neumann boundary conditions, i.e. vanishing of the derivatives, at $x=-\infty$ and
$x=+\infty$. One can rewrite the equation as
\begin{eqnarray}&&
\frac{d\phi}{dx}=\frac{T(\phi)[2+\cosh 2\phi][2T^4(\phi)\sinh 2\phi-C_1]}{F(\lambda)\cosh\phi [2T^4(\phi)\sinh 2\phi+C_1/3]} \ , \label{eq105}
\end{eqnarray}
with $T(\phi)$ defined by Eq.~(\ref{eq101}). We find that $v=\tanh \phi$ obeys
\begin{eqnarray}&&
\frac{dv}{dx}=\frac{3^{3/4}\zeta C_2^{1/4}[\zeta-v]^{1/4}[3-v^2][v-v_1][v-v_2]\sqrt{1-v^2}}{F(\lambda)v^{1/4}[\zeta(5-v^2)-4v]} \ , \label{eq125}
\end{eqnarray}
where $\zeta\equiv C_1/C_2$ and $v_{1, 2}$ are the roots of $\zeta(1+3v^2)-4v=0$,
\begin{eqnarray}&&
v_{1,2}=\frac{2\pm\sqrt{4-3\zeta^2}}{3\zeta}.
\end{eqnarray}
Note that $v_1=1/3v_2$, cf. Eq.~(\ref{idshock}). The viscous counterpart of the ideal hydrodynamics shock considered in the first subsection
corresponds to the range $1\leq \zeta<2/\sqrt{3}$ where $v_{1, 2}$ are real and $1/3\leq v_{1,2}\leq 1$.

We first consider the case $v(-\infty)<v(+\infty)$ and search for the solution of Eq.~(\ref{eq125}) satisfying
\begin{eqnarray}&&
v(-\infty)=v_2=\frac{2-\sqrt{4-3\zeta^2}}{3\zeta},\ \ \ v(+\infty)=v_1=\frac{2+\sqrt{4-3\zeta^2}}{3\zeta}.
\end{eqnarray}
Such a solution must have a positive derivative and
values close to $v_2$ in the range $v_2<v<v_1$ in some region of space. However, Eq.~(\ref{eq125}) dictates that at least
close to $v_{1, 2}$ the sign of $dv/dx$ is determined by $[v-v_1][v-v_2]$ which is negative at $v_2<v<v_1$. This follows by
observing that the factor $\zeta(5-v^2)-4v$ in the denominator of Eq.~(\ref{eq125}) obeys $\zeta(5-v^2)-4v=3\zeta[v-v_1][v-v_2]+4\zeta[1-v^2]$.
Thus it is not possible to construct the solutions with $v(-\infty)<v(+\infty)$.

In contrast, the solutions with $v(-\infty)>v(+\infty)$
for which
\begin{eqnarray}&&
v(-\infty)=v_2=\frac{2+\sqrt{4-3\zeta^2}}{3\zeta},\ \ \ v(+\infty)=v_1=\frac{2-\sqrt{4-3\zeta^2}}{3\zeta}.
\end{eqnarray}
are easily constructed. Noting that
$\zeta(5-v^2)-4v$ is always positive for $v_2<v<v_1$ at $\zeta>1$ we conclude from Eq.~(\ref{eq125}) that $v(x)$ decreases monotonically
from $[2+\sqrt{4-3\zeta^2}]/3\zeta$ at $x=-\infty$ to $[2-\sqrt{4-3\zeta^2}]/3\zeta$ at $x=+\infty$. Here we use that it takes the solution
of an equation of the type $dv/dt=H(v)$ infinite "time" $t$ to reach a zero of an analytic function $H(\phi)$. The implicit form of the
solution is given by
\begin{eqnarray}&&
x-x_0=\frac{F(\lambda)}{3^{3/4}\zeta C_2^{1/4}}\int_{2/3\zeta}^v \frac{v'^{1/4}[\zeta(5-v'^2)-4v']dv'}{[\zeta-v']^{1/4}[3-v'^2][v'-[2+\sqrt{4-3\zeta^2}]/3\zeta][v'-[2-\sqrt{4-3\zeta^2}]/3\zeta]\sqrt{1-v'^2}},
\label{implicit}
\end{eqnarray}
where the constant $x_0$ is the point at which $v(x)$ attains its middle value $2/3\zeta$ (naturally the solution is translation-invariant).
In the following we set $x_0=0$. Note that $v(x, C_1, C_2)=v(C_1/C_2, C_2^{1/4}x)$. The dependence of temperature on $x$ is
determined by $T^4(x)=C_2[\zeta-v(x)]/3v(x)$. One observes that the necessary condition that the resulting expression
for $T^4$ is positive is satisfied by $\zeta>1$ and $v<1$. The ideal hydrodynamics shock is recovered in the formal limit
of zero viscosity, $F(\lambda) \rightarrow 0$.

Thus the viscosity selects among the ideal hydrodynamics shocks only those for which the inflow velocity $v(-\infty)$ is
larger than the outflow velocity $v(+\infty)$. This has a simple meaning: the dropping of velocity corresponds to a dissipative action of
the viscosity (friction), while velocity increase would qualitatively mean an accelerating action of the friction. The selected flows
start as supersonic and end as subsonic.

It remains to consider the condition under which the above solution has a small Knudsen number to justify neglecting the
higher derivative terms in $T^{0x}$ and $T^{xx}$. We notice that generally $F(\lambda)$ is of the same order as $G(\lambda)$ appearing
in the expression (\ref{eq112}) for $l_{cor}$, see \cite{FBB,Fouxon:2008tb}. We define the local Knudsen number by
\begin{eqnarray}&&
Kn=l_{cor}\left|\frac{v'}{v}\right|=
\frac{3\zeta [3-v^2]|[v-v_1][v-v_2]|\sqrt{1-v^2}}{v[\zeta(5-v^2)-4v]}, \label{eq114}
\end{eqnarray}
where we used Eqs.~(\ref{eq112}) and (\ref{eq125}). The viscous shock solution above is valid if $Kn\ll 1$ uniformly.
At fixed $\zeta$, by choosing a sufficiently small $C_2$ one may achieve an arbitrarily large scale of variations of $v$ in space, due to
$v(x, C_1, C_2)=v(C_1/C_2, C_2^{1/4}x)$. This, however, does not lead to a solution with a small Knudsen number, since the correlation length also
increases under such scaling of $C_i$. This feature is taken into account by Eq.~(\ref{eq114}), that depends only on the ratio $C_1/C_2$.

To check that $Kn\ll 1$, it is enough to substitute in the above equation
the middle value $2/3\zeta$ of $v$ which produces a term of order $4-3\zeta^2$ ($\zeta$ belongs to the range $[1, 2/\sqrt{3}]$.
It follows that $Kn\ll 1$ is satisfied only for $\zeta$ close to $2/\sqrt{3}$ (a more precise condition is derived below). For such $\zeta$, the inflow velocity $v(-\infty)$ is close
to the speed of sound $1/\sqrt{3}$, and the outflow obeys $v(+\infty)\approx v(-\infty)$, cf. remarks to Eq. (\ref{idshock}) . The Knudsen
number is small just because the total variation of $v$ is small. In other words, for this solution the requirement of locally small variations
of fields, implies also that the global variation is small.

\subsection{The viscous shock solution at a small Knudsen number}

In the small Knudsen number domain $\zeta=2/\sqrt{3}-\epsilon$ with $\epsilon\ll 1$, the velocity changes in the interval
$[1/\sqrt{3}-\delta, 1/\sqrt{3}+\delta]$ where $\delta=\epsilon^{1/2}3^{-1/4}$. Using that the total variation of the solution is
small, one can obtain more explicit expressions in this case. To leading order in $\delta$ Eq.~(\ref{eq125}) becomes
\begin{eqnarray}&&
\frac{d{\tilde v}}{dx}=\frac{3^{1/4}\sqrt{2}C_2^{1/4}[{\tilde v}-\delta][{\tilde v}+\delta]}{F(\lambda)} \ , \label{eq145}
\end{eqnarray}
where ${\tilde v}=v-1/\sqrt{3}$ is the velocity deviation from the speed of sound. The solution of the above equation gives the
well-known kink solution, which can also be obtained by expanding the integral (\ref{implicit}),
\begin{eqnarray}&&
v=\frac{1}{\sqrt{3}}-\delta\tanh\left[\frac{x}{L}\right],\ \ L=\frac{F(\lambda)}{3^{1/4}\sqrt{2}C_2^{1/4}\delta}\sim \frac{l_{cor}}{\delta}, \label{eq130}
\end{eqnarray}
where we used that $T\approx [C_2/3]^{1/4}$ holds for the solution. One has
\begin{eqnarray}&&
T^4\approx \frac{C_2}{3}+\frac{2C_2\delta}{\sqrt{3}}\tanh\left[\frac{x}{L}\right].
\end{eqnarray}
We see that the Knudsen number $Kn=l_{cor}/L\sim \delta$ is small at $\delta \ll 1$. Note, that the Knudsen number definition (\ref{eq114})
would give incorrectly $Kn\sim \delta^2$ - the reason is that $v(x)$ has a special form where $v'(x)$ is determined by the
small correction ${\tilde v}$ to the main constant part $1/\sqrt{3}$.

One can exclude the constant component in $v(x)$ by passing to the frame moving at the speed of sound, where the solution takes the form of a kink
propagating at the sound speed,
\begin{eqnarray}&&
\frac{v}{c}=-\frac{3\delta}{2}\tanh\left[\sqrt{\frac{3}{2}}\left(\frac{x+ct/\sqrt{3}}{L}\right)\right],\ \ T^4\approx \frac{C_2}{3}+\frac{2C_2\delta}{\sqrt{3}}\tanh\left[\sqrt{\frac{3}{2}}\left(\frac{x+ct/\sqrt{3}}{L}\right)\right],
\label{solv}
\end{eqnarray}
in the leading order in $\delta$, where we restored $c$ in $v$. One sees that the solution is a near equilibrium state.
In the limit $c\to\infty$ the motion gets degenerated to $v=const$ in accord with the fact that the only incompressible one-dimensional
flow is a constant one.

\subsection{The shock solution at a Knudsen number of order unity}

We saw above that if the shock solution is such that the inflow velocity $v(-\infty)$ differs significantly from the speed of
sound then there are regions in the flow where $Kn\sim 1$ (this follows from self-consistency: if $Kn\ll 1$ then viscous hydrodynamics
should work well). The solution of the viscous hydrodynamics then does not describe the solutions correctly
in those regions though it may give a good idea of the solution there. To find an exact solution one should account for higher order
derivatives in the constitutive relation for $T^{\mu\nu}$ (for instance, the second derivative terms were discussed in \cite{Son,HaackYarom}).
The above however does not undermine the utility of the ideal hydrodynamics shock which is still a good description of the solution
far from the transition region near $x=0$.

We show this point explicitly. We consider a non-equilibrium steady state of a CFT fluid which is sustained by the fluxes of energy
and $x-$component of momentum from infinity. In this state the mean value of $T^{\mu\nu}$ depends only on $x$ and Eqs.~(\ref{basic110})
are satisfied irrespective of the validity of the hydrodynamic constitutive relations. Now, far from the transition region the scale
of variations of the mean value of the stress-energy tensor becomes large, the state is locally very close to equilibrium and the
ideal hydrodynamics approximation $T_{\mu\nu}= T^4\left[\eta_{\mu\nu}+4u_{\mu}u_{\nu}\right]$ works with increasing degree of accuracy.
In other words, far from the transition region the state looks locally closer and closer to the global equilibrium
(the information that the local equilibrium state is not the global one becomes further separated in space). As a result, the ideal
hydrodynamics shock solution allows to establish correctly at all orders the connection between the asymptotic values of  $v(-\infty)$, $T(-\infty)$ and $v(+\infty)$, $T(+\infty)$, and thus also of $T^{\mu\nu}(-\infty)$ and $T^{\mu\nu}(+\infty)$. As far as the intermediate region of the transition is concerned, it is described by the solution in the previous subsection for the inflow velocity close to the speed of sound $1/\sqrt{3}$.
When the inflow velocity $1/\sqrt{3}<v(-\infty)<1$ is not close to $1/\sqrt{3}$ the description of the transition region demands the account
of the higher order derivatives. We do not expect, however, that the higher derivative terms will lead to a further selection of the ideal
hydrodynamic shocks beyond the one imposed by the viscosity.

\subsection{How is it possible to have $Kn\ll 1$, while the viscous term still being important?}

It is natural to ask why is it possible to construct a solution where the viscous term, whose effect is generally measured by the
Knudsen number, is important while higher order derivative terms, whose effect is measured by the same parameter, are not.
The reason being
the presence of an additional dimensionless parameter in the equations, $v/c$ (here we restore $c$ in the relations). As long as
we consider relativistic motions with $v\sim c$, the viscous to ideal terms ratio is governed by the same parameter $Kn$ as the
ratio of higher order derivative terms to the viscous one. However, as we pass to motions with $v\ll c$ the viscous to ideal terms ratio
involves an additional small parameter $v/c$. In the case of properly defined non-relativistic motions obeying the incompressible Navier-Stokes equations the dimensionless parameter governing the viscous to ideal terms ratio is the inverse of the
Reynolds number $Re$
\begin{equation}
Re\sim vL/\nu \ ,
\end{equation}
where $v$ is the characteristic velocity, $L$ is the characteristic scale and $\nu$ is the kinematic viscosity, see Eq.~(\ref{NS}). Thus,
one can have a situation, where in some regions of space the Reynolds number is not large and viscosity is important, while the Knudsen
number is still small everywhere. As long as the viscosity prevents the generation of smaller scales by the dynamics (as it does at least in known cases), one obtains a global solution of the hydrodynamics with viscous term important but higher order derivatives not.
In other words, while viscosity is typically a singular perturbation producing order one effects in the solution \cite{Feynman}, the higher
order terms are often regular perturbations.

The relevance of the above discussion to the viscous shock solution considered before, is that at small Knudsen number, $v(x) \ll c$ in (\ref{solv}).
Thus, in the frame moving at the speed of sound one deals with
the situation of small $v/c$ and the above picture holds. Let us stress again, however, that this solution has no well-defined non-relativistic
limit as the shock propagates at a speed of order $c$.

Note also, that here and below we use the notion of a local Reynolds (Knudsen) number, which is the Reynolds (Knudsen) number defined via the characteristic values of the scale and velocity in the considered region of space \cite{Frisch}.

\section{Gravity and fluid dynamics}
\label{gravity}

It has been shown in \cite{Bhattacharyya:2008jc}, that the four-dimensional CFT hydrodynamics equations are the same as
the equations describing the evolution
of long-scale perturbations of the five-dimensional black brane. In fact, the derivation of this result parallels
exactly the conventional derivation of hydrodynamics equations from the Boltzmann equation. Let us show this in the simplest case of a
non-relativistic monoatomic gas.

The Boltzmann equation for a non-relativistic monoatomic gas  reads \cite{Landau10}
\begin{eqnarray}&&
\partial_t f+\bm V\nabla_{\bm x}f=St f \ , \label{Boltzmann}
\end{eqnarray}
where $f(\bm x, \bm p, t)$  is the density function in a single particle phase space,
$\bm V=\bm p/m$ where  $m$ is the particles mass, and $St f$ is the collision integral, which is quadratic in $f$.
Equation (\ref{Boltzmann}) has a steady-state solution of a Maxwell distribution form
\begin{eqnarray}&&
f(\bm x, \bm p)=\frac{N}{[2\pi m k_BT]^{3/2}}
\exp\left[-\frac{m\left(\bm V-\bm v\right)^2}
{2k_BT}\right] \ , \label{Maxwell000}
\end{eqnarray}
where $N$ is the mean concentration of particles, $T$ is the temperature and
$\bm v$ is an arbitrary velocity present due to the Galilean invariance of Eq.~(\ref{Boltzmann}).

The
hydrodynamics equations arise when solving the Boltzmann equation by the method of variation of constants.
Consider a solution in the form
\begin{eqnarray}&&
f(\bm x, \bm p, t)=f_0(\bm x, \bm p, t)+\delta f(\bm x, \bm p, t),\ \
f_0(\bm x, \bm p, t)=\frac{N(\bm x, t)}{[2\pi m k_BT(\bm x, t)]^{3/2}}
\exp\left[-\frac{m\left[\bm V-\bm v(\bm x, t)\right]^2}
{2k_BT(\bm x, t)}\right], \label{ansatz}
\end{eqnarray}
where $\delta f\ll f_0$. This describes the local thermal equilibrium in the gas.
The construction of a solution to the Boltzmann equation where $f_0$ is the zeroth order approximation, produces
a series in the Knudsen number $Kn$, i.e. the ratio of the mean free path to the scale of spatial variations of the fields.
Thus,
$\delta f\sim Kn f_0$, and the hydrodynamics equations
governing the evolution of $N(\bm x, t)$, $T(\bm x, t)$ and $\bm v(\bm x, t)$ result as the conditions of the series constructibility.

Analogous derivation of hydrodynamics from the (relativistic) Boltzmann equation can be performed for a CFT. This becomes
possible in the limit of the weak coupling where quasiparticle excitations are well-defined (see e.g. \cite{Son}). In contrast,
at a strong or order unity coupling the Boltzmann equation does not apply. What this means normally is, that for non weakly coupled
fluids the hydrodynamics equations are valid but cannot be derived from a microscopic description. The CFT fluids, however,
bring a surprise.

Consider the derivation of the CFT hydrodynamics at strong coupling from the equations of gravity.
The analogue of Eq.~(\ref{Boltzmann}) is given by the five-dimensional Einstein equations with negative cosmological constant
\begin{eqnarray}&&
R_{mn}+4g_{mn}=0,\ \ R=-20 \ . \label{u4}
\end{eqnarray}
These equations have a particular "thermal equilibrium" solution - the boosted black brane
\begin{eqnarray}&&
ds^2=-2u_{\mu}dx^{\mu}dr+\pi^4 T^4 r^{-2}u_{\mu}u_{\nu}dx^{\mu}dx^{\nu}+r^2\eta_{\mu\nu}dx^{\mu}dx^{\nu} \ ,\label{u3}
\end{eqnarray}
where the temperature $T$ is constant. The presence of an arbitrary constant vector $u^{\mu}$ in the above solution is due to Lorentz
invariance, much like the presence of $\bm v$ in Eq.~(\ref{Maxwell000}) due to Galilean invariance. As in Eq.~(\ref{ansatz}), one looks
for a solution of the Einstein equation by the method of variation of constants using the ansatz
\begin{eqnarray}&&
\!\!\!\!\!\!\!\!\!\!\!\!\!\!g_{mn}=(g_0)_{mn}+\delta g_{mn},\ \ (g_0)_{mn}dy^{m}dy^{n}=-2u_{\mu}(x^{\alpha})dx^{\mu}dr+\pi^4 T^4(x^{\alpha}) r^{-2}u_{\mu}(x^{\alpha})u_{\nu}(x^{\alpha})dx^{\mu}dx^{\nu}+r^2\eta_{\mu\nu}dx^{\mu}dx^{\nu},\label{gravityansatz}
\end{eqnarray}
where $y=(x^{\mu}, r)$. As in the Boltzmann equation, the condition of constructibility of the series
solution produces equations for $u^{\mu}(x^{\alpha})$ and $T(x^{\alpha})$.  These equations
are the hydrodynamics equations for a certain class of CFTs in the limit of infinite coupling \cite{Bhattacharyya:2008jc}.
As far as the hydrodynamic equations
are concerned the infinite coupling means only a particular value of constants in the relations. For example, for dissipative
hydrodynamics considered in the previous sections infinite coupling means taking instead of $F(\lambda)$ its appropriate $\lambda\to\infty$
limit, $F=1/\pi$. Note, that at the level of the ideal hydrodynamic equations
there is no difference at all between different CFTs - the equations are the same. The resulting series for
$g_{mn}$ is the series in the Knudsen number of the boundary CFT hydrodynamics.

Thus equations of gravity play for the hydrodynamic equations of a strongly coupled CFTs much the same role as the Boltzmann
equation plays at a weak coupling. It is natural to inquire whether the equations of the CFT hydrodynamics allow for a microscopic
derivation also at an arbitrary intermediate coupling (say from the equations of gravity with higher order curvature terms in the action).
This important question is beyond the scope of this work and is a subject for the future research.

We conclude from the above that the solutions to the CFT hydrodynamics at an infinite coupling describe not only the large-scale
motions of the theory, but also the solutions to gravity corresponding to Eq.~(\ref{gravityansatz}).
We may now use the general results for the CFT
hydrodynamics, obtained in the previous section and valid in particular for the hydrodynamics of the infinitely coupled CFT resulting
from Eqs.~(\ref{u4}), in order to gain some insight into gravity
dynamics. If we consider initial conditions for gravity, that correspond to the initial conditions
for hydrodynamics with $v\sim c$ and $Kn\ll 1$ everywhere, then for some limited amount of time we may use ideal
hydrodynamics to determine the metric evolution. The ideal hydrodynamic flow is compressible, and it leads to the formation
of discontinuities (and maybe other, yet unknown types of singularities) in a finite time. In particular, for "one-dimensional" initial conditions, one may use the exact solutions obtained in the previous sections. Then, ideal hydrodynamics predicts that in a finite time, regions with large metric curvature will appear. They correspond to large values of gradients of the hydrodynamics fields. Eventually, in these regions
higher order derivative terms will become important for the evolution of $u^{\mu}$ and $T$ and (expectedly) stop the growth of the curvature.
One expects $Kn\sim 1$ and $\delta g\sim g_0$ in these regions.

\subsection{The importance of the solutions to non-relativistic hydrodynamics for the construction of global solutions to gravity}

As already explained in the last subsection of section \ref{shocks} the consideration of non-relativistic solutions with $v\ll c$
allows one to construct global solutions of hydrodynamics with $Kn\ll 1$ everywhere. Such solutions are especially important from
the viewpoint of the construction of globally valid approximations to the solutions of gravity. Indeed, for them one has
the explicit form $g\approx g_0$ of the approximate solution to the Einstein equations (\ref{u4}). This is quite similar to the
situation for non-relativistic classical gas where the local thermal equilibrium $f\approx f_0$ holds even though viscosity
is important for the evolution of the local parameters of equilibrium $T$, $N$ and $\bm v$, cf. \cite{Landau10}.

The explicit form of the approximate solution to the Einstein equations resulting from the solution to the non-relativistic incompressible
Navier-Stokes equations was summarized in Eqs.~(\ref{gravityansatz01})-(\ref{gravityansatz02}).

\subsection{Global solution for gravity from a small Knudsen number viscous shock}

As an example of a global solution to gravity obtained from hydrodynamics we consider the five-dimensional asymptotically AdS gravity solution that corresponds to the viscous shock. Using the results of the previous section we find that the metric $g_{vs}$ defined by
\begin{eqnarray}&&
(g_{vs})_{mn}dy^{m}dy^{n}=\sqrt{6}\left[1-\frac{\sqrt{3}\delta\tanh(x/L)}{2}\right]dtdr-\sqrt{2}\left[1-\frac{3\sqrt{3}\delta\tanh(x/L)}{2}
\right]dxdr+r^2\eta_{\mu\nu}dx^{\mu}dx^{\nu}\nonumber\\&&+\frac{\pi^4C_2}{3r^{2}}\left[1+2\sqrt{3}\delta\tanh\left(\frac{x}{L}\right)\right]
\Biggl[\left[1-\sqrt{3}\delta\tanh\left(\frac{x}{L}\right)\right] \frac{3dt^2}{2} -\sqrt{3}\left[1-2\sqrt{3}\delta\tanh\left(\frac{x}{L}\right)\right]dxdt
\nonumber\\&&+\left[1-3\sqrt{3}\delta\tanh\left(\frac{x}{L}\right)\right] \frac{dx^2}{2}\Biggr],\ \ L=\frac{1}{3^{1/4}\sqrt{2}\pi C_2^{1/4}\delta},
\label{shugar}
\end{eqnarray}
is a solution of the Einstein equations to order $\delta^2\ll 1$ (the constant $C_2>0$ is arbitrary), and indeed, an explicit calculation verifies that.

Above we used that at small
$\delta$ one has $u_0=-[1-v^2]^{-1/2}
\approx -\sqrt{3/2}[1-\sqrt{3}\delta\tanh(x/L)/2]$ and $u_x=v[1-v^2]^{-1/2}\approx 1/\sqrt{2}[1-3\sqrt{3}\delta\tanh(x/L)/2]$,
where $v(x)$ is given by Eq.~(\ref{eq130}).
As we saw in the previous section the requirement $Kn\ll 1$ imposes that the total variation of $v$ is small and the above solution describes
a near equilibrium state.

The above stationary solution corresponds to the solution of the Einstein equations
with the
boundary condition imposed on the boundary stress tensor \cite{Balasubramanian:1999re}
\begin{equation}
T^{\mu}_\nu = -2\lim_{r \to \infty}
r^4 \left( K^{\mu}_{\nu} -\delta^\mu_\nu \right) \ ,
\end{equation}
where $K_{\mu\nu}$ is the extrinsic curvature tensor to the surface at fixed $r$ and $T^{\mu\nu}$ is the solution of the hydrodynamics equations.
If one used the ideal hydrodynamics approximation, where the shock is discontinuous one would obtain
a domain wall solution for gravity with no $\delta-$function sources at the wall. At small $Kn$, the solution (\ref{shugar}) allows to resolve the inner structure of the wall. At $Kn\sim 1$, the resolution of the wall structure requires the inclusion of the higher order derivatives in the analysis. However,
we stress again that the ideal shock solution provides a correct description of the domain wall solution far from
wall at any $Kn$.
Finally, we note that it is straightforward to generalize this domain wall solution of gravity by applying special conformal transformations to the solution.

\medskip\medskip
\noindent{\bf Acknowledgements:}
This work is supported in part by the Israeli Science Foundation center of excellence,
by the Deutsch-Israelische Projektkooperation, by the the US-Israel binational science foundation and
by the European Network.

\end{document}